# Simulation Study on a New Peer Review Approach

Albert Steppi[1], Jinchan Qu[1], Minjing Tao[1], Tingting Zhao[2,3], Xiaodong Pang[3], Jinfeng Zhang[1,3*]
[1]Department of Statistics, Florida State University
[2]Department of Geography, Florida State University
[3]Pevals LLC, Tallahassee, FL

*Abstract—* The increasing volume of scientific publications and grant proposals has generated an unprecedentedly high workload to scientific communities. Consequently, review quality has been decreasing and review outcomes have become less correlated with the real merits of the papers and proposals. A novel distributed peer review (DPR) approach has recently been proposed to address these issues. The new approach assigns principle investigators (PIs) who submitted proposals (or papers) to the same program as reviewers. Each PI reviews and ranks a small number (such as seven) of other PIs' proposals. The individual rankings are then used to estimate a global ranking of all proposals using the Modified Borda Count (MBC). In this study, we perform simulation studies to investigate several parameters important for the decision making when adopting this new approach. We also propose a new method called Concordance Index-based Global Ranking (CIGR) to estimate global ranking from individual rankings. An efficient simulated annealing algorithm is designed to search the optimal Concordance Index (CI). Moreover, we design a new balanced review assignment procedure, which can result in significantly better performance for both MBC and CIGR methods. We found that CIGR performs better than MBC when the review quality is relatively high. As review quality and review difficulty are tightly correlated, we constructed a boundary in the space of review quality vs review difficulty that separates the CIGR-superior and MBC-superior regions. Finally, we propose a multi-stage DPR strategy based on CIGR, which has the potential to substantially improve the overall review performance while reducing the review workload.

*Keywords:* peer review, distributed peer review, concordance index-based global ranking, CIGR, Modified Borda Count,

**Abbreviations and Notations**

DPR: Distributed peer review

CI: Concordance Index

CIGR: Concordance Index-based Global Ranking

MBC: Modified Borda Count

$SD_s$: Standard Deviation of the underlying true proposal/paper scores

$E_r$: Error of a reviewer's scores. This variable is used to simulate the accuracy of the scores given by a reviewer compared to the true scores. Smaller errors correspond to higher quality of reviews. $E_{r,i}$ is the average error of reviews for the reviewer $i$.

$B_r$: Bias of a reviewer. This variable is used to simulate the inherent bias a PI may have towards all the proposals/papers he/she will evaluate. For example, someone tends to give higher (or lower) scores for all the proposals/papers assigned to him/her.

$n_p$: Number of PIs in the pool

$n_r$: Number of times a proposal will be reviewed by PIs, which is the same as the number of proposals assigned to a PI

I. INTRODUCTION

Scholarly research assessment plays indispensable roles in many aspects of the scientific research enterprise. Peer-reviewed scientific articles are usually evaluated by a small number of experts in the relevant research areas. Grant proposals are judged by panels of experts of varying sizes. Peer review has received a lot of attention since its outcome can dramatically affect the career paths of scientists whose papers or grant proposals are subject to this process [1-5]. Compared to the editor-inclusive referee system, peer review from a panel of outside experts was considered as a more trustworthy means to evaluate the merit of a research paper or proposal. In recent years, as academia is experiencing an exponential growth in the number of scientific manuscript and grant proposal submissions, increased review workload has been piled upon reviewers. The burden placed on reviewers consequently affects the review quality, and can even cause stress that prevents reviewers from taking the process seriously. A series of articles has questioned the benefit of peer review [1, 6-8]. Suggestions to improve the review quality have also been proposed [7-12], such as training reviewers and providing more strict guidelines.

To meet the dramatic growth in scientific research submissions, a distributed peer review (DPR) process was proposed by Merrifield and Saari [13] to reduce the burden on the reviewer community, lower the costs associated with the review process, while also potentially improve the quality of reviews. The authors considered the common setting of a pool of applicants who have submitted proposals to a program. Each applicant was assigned several proposals that had been submitted by other applicants in the pool. Applicants each acted as a reviewer to evaluate and rank the proposals assigned to them. The partial rankings submitted by all applicants were then used to generate a global ranking for all of the proposals. Modified Borda Count (MBC), a voting rule that assigns points to each proposal according to its position in the input ranking lists, was used to infer the global ranking. The MBC score for a proposal was calculated by summing up all the points from its reviewers, which is then normalized by the maximum possible points a proposal can receive. An aggregate ranking was then obtained based on the values of MBC scores. Meanwhile, incentives were designed to encourage reviewers to work out a ranking in line with the group consensus. A PI's proposal would be ranked by its inherent merits and then adjusted by how the PI's ranking compared to the initially inferred global ranking.

DPR was later tested by the National Science Foundation (NSF) in a pilot study in 2013 [14]. PI group sizes of 25 to 40 were used, and 7 proposals were assigned to each PI to review. Naghizadeh and Liu [15] provided a detailed discussion of the strengths and issues of this peer-review system in the NSF pilot test, focusing on the effects of incentives for review quality and the potential for collusion between PIs. They made a conclusion in favor of DPR after comparing its benefits and weaknesses. DPR has been adopted by the Gemini Observatory in their Fast Turnaround grant program [16]. In this program, each review cycle takes only one month. DPR was also used recently by the US Department of Agriculture (USDA) as a pilot in one of their grant programs [17]. The following advantages were mentioned in the USDA pilot program announcement: DPR "(1) minimizes the time between Request for Application (RFA) closing and applicant feedback; (2) places the burden of peer review proportionally on those who burden the review system; (3) incentivizes unbiased and timely reviews that strive for consensus; (4) increases applicant feedback; and (5) reduces costs, facility resources and staff time."

Given the advantages of DPR, current and future practitioners need guidance on several key issues when adopting DPR in their review tasks. For example, what is the optimal size of the reviewer pool? What is the optimal number of reviews to be assigned to each reviewer to permit balance between high review quality and low workload? The quality of DPR depends on the quality of reviews provided by the applicants. For scientific proposals, the PIs who submit proposals are considered to be highly qualified.

This may not be true for other applications. What is the minimum review quality of the applicants for the method to achieve a certain performance threshold? These are important parameters one would like to know before applying DPR to their review tasks.

Computation of MBC depends on the random assignment of proposals to reviewers. However, a well-designed assignment combined with appropriate algorithms may achieve better ranking performance. For example, if prior knowledge of approximate overall rankings or scores is known, assigning pairs of proposals/papers of similar quality to the same reviewer may help better distinguish between these pairs, which in turn helps to potentially improve the accuracy of the global ranking. In such cases, MBC will not give useful global rankings because all the proposals except the top- and bottom-ranked ones will have similar MBC values. Alternative metrics to MBC suitable for non-random assignments will allow researchers to explore new algorithms which may significantly improve the performance of DPR.

In this study we propose a new approach, called concordance index-based global ranking (CIGR) for inferring global rankings from individual rankings. The idea is to search for a global ranking that maximizes agreement with the input list of partial rankings. We choose to maximize the concordance index (CI), which is defined as the proportion of the pairwise counts by PIs that are ranked consistently in the candidate global ranking. It is similar to the Kendall distance used in Kemeny's method [18], which counts the number of pairwise disagreements between the global ranking and individual rankings. Finding a global ranking with maximum CI with respect to the individual rankings is computationally intractable [18-21]. We use a Markov Chain Monte Carlo (MCMC) method with simulated annealing to find near optimal rankings. The method then aggregates over near optimal rankings using MBC, achieving greater stability than a direct MCMC approach.

When reviewers evaluate proposals they can either submit ranks or numerical scores, the latter of which can be used to derive ranks. Scores carry more information, which can be helpful when inferring a global ranking for all the proposals. However, scoring proposals can take more effort compared to ranking them, especially when the review load is low (for example less than 5 or 7 reviews per reviewer). In this study we focus on evaluations where only ranks are provided since reducing reviewers' workload is one of the considerations driving attempts to change the current review system. Rank-based approaches can also be generalized more easily to other related applications, such as paper evaluations, where reviewers are not usually required to provide scores.

In this work, we perform simulation studies to investigate the aforementioned questions important for decision making when using DPR. We study the effects of several parameters related to the performance of DPR, including the size of the reviewer pool, the number of proposals/papers assigned to each reviewer, the distribution of the underlying true scores, the bias of the reviewers, and the review quality of the reviewers. We compare the performance of MBC-based approach and CIGR on various conditions. Interestingly, we find that neither MBC nor CIGR outperforms the other in all scenarios. We identify the situations where MBC or CIGR would perform better to provide guidance for DPR applications where both approaches can be used.

There are many ways one can assign proposals to each reviewer. Given the framework of random assignments, is it possible to choose assignments in a manner that improves the quality of generated global rankings? We investigate the effect of a more balanced assignment design, where the number of times that each pair of proposals is assigned together to a reviewer is made as uniform as possible. We find that this balanced assignment design can significantly improve global ranking performance for both MBC and CIGR. It also makes use of an MCMC algorithm with simulated annealing, in this case, through maximizing the entropy of an assignment.

The rest of the paper is organized as follows. Section II describes the details of the simulations and the methods for obtaining global rankings, including our new method CIGR. Evaluations of performance and the new balanced review assignment design are provided in Section III. We conclude the paper and discuss future work in Section IV. The details of the simulated annealing algorithms for optimizing CI and generating balanced assignments are provided in the Appendix.

## II. METHODS

### 1) Simulation setup

Here we describe our simulations of the proposal review process. We will use the correponding terms in this section and the rest of the paper. The applicants, or PIs, will also be referred to as reviewers. Each applicant submits one proposal to the system.

Given the number of applicants or PIs, $n_p$, we first simulate the scores of their proposals using a truncated normal with mean 50, lower and upper bounds of 0 and 100 respectively, and a standard $SD_s$. The proposals are then ranked according to their true scores to generate a true global ranking. In the process, each reviewer will review and score $n_r$ proposals that have been assigned to them, these scores will then be used to generate a ranking of these $n_r$ proposals. The scores given to the proposals were not used in inferring global rankings for the reasons mentioned in the introduction. We use a normal distribution with mean, $\mu_i$, and standard deviation, $\sigma_i$, to model the $i$-th reviewer's review quality. The value sampled from this normal distribution, $N(\mu_i, \sigma_i)$, will be added to the true score of a proposal to generate the reviewer's score for this proposal. Some applicants tend to give all the proposals higher scores, while others tend to give proposals lower scores. The mean parameter, $\mu_i$, represents this inherent bias. The standard deviation, $\sigma_i$, models how much their scores deviate from the true scores, with larger standard deviations corresponding to larger errors. These two parameters are kept the same for all the proposals a reviewer will evaluate. The values of $\mu_i$ are sampled from normal distributions with mean 0 and several different standard deviation values (see details in Table 1) to model different magnitudes of reviewers' bias. The values of $\sigma_i$ are sampled from chi-squared distributions with several different degrees of freedom. We refer to $\mu_i$ as the bias of a reviewer ($B_r$), and $\sigma_i$ as the average error of the reviewer's scores ($E_r$). In Table 1, we list all the parameters that are used in the simulation study and the values they take on.

TABLE 1. The parameters in the simulation study and their possible values.

| Parameter names | What the parameter model is | Values simulated |
|---|---|---|
| **Number of PIs, $n_p$** | How the size of the applicant pool affects the ranking accuracy | 10 - 200 |
| **Number of proposals each applicant review, $n_r$** | How the number of proposals each applicant reviews affects the global ranking accuracy | 3, 5, 7, 9, 11, 13, 15 |
| **The standard deviation of proposal scores, $SD_s$** | How the distribution of the proposal scores affects the ranking accuracy | (0, 30] |
| **The (mean, standard deviation) for bias of a reviewer, $B_r$** | The magnitudes of the reviewers' tendency to score all proposals higher or lower than their true scores | (0, 1) to (0, 20) |
| **The degrees of freedom limit (lower, upper) for the chi-squared distribution for the average error of the reviewer's scores, $E_r$** | The error of PI's scores compared to the true scores. | (5, 20] |

To evaluate the performance of a ranking method, we use two criteria. One is to measure the ranking accuracy for all proposals by using Concordance Index (CI), which calculates the proportions of proposal pairs given by PIs that are ordered consistently in the candidate global ranking. We refer to this criterion as ranking accuracy. The other criterion is to calculate the proportion of proposals that can be correctly selected as lying in a top percentage of proposals. For example, we may want to know what fraction of the top 10% of proposals in the true ranking is also in the top 10% of proposals in the inferred ranking. In this criterion, there is no focus on the ranking accuracy within the top 10% of proposals. This criterion is especially useful in the scenario of selecting the winners/awardees. It is very common that what a funding agency (or a journal) needs is to filter out 80% of the proposals (or papers) and only select the top 20% of proposals. The relative rankings are less of interest. We refer to this criterion as filtering accuracy. In the evaluations we check the performance for the top 20% of proposal and refer to this criterion as $T_{0.2}$.

*2) Modified Borda Count (MBC)*

MBC allows each PI to submit an ordered list and points are given to the proposals to indicate their relative positions in the list. For instance, to rank $n_r$ proposals, the points are assigned in the range (0, $n_r$-1) with 0 corresponding to the worst ranked proposal and $n_r$-1 to the best ranked proposal. A proposal's MBC score is then the sum of all $n_r$ values (given by $n_r$ reviewers) divided by $n_r(n_r$-1). Though ties in the ordered list are not encouraged, proposals are allowed to apportion the points with the same weight, such as a 0.5 scale on each of the two tied proposals, or 0.33 on each of the three tied proposals, etc. [14]. The total amount of points given to $n_r$ proposals remains the same.

Note that an unfair assignment usually results in MBC not performing its best. It possible a worse proposal can be ranked ahead of a better one if it has more pairwise comparisons with poorer proposals.

*3) Concordance Index based Global Ranking (CIGR)*

We propose a new method CIGR that is based on maximizing the agreement between candidate global rankings and individual rankings provided by reviewers. This agreement is measured by using the Concordance Index (CI). This is similar to an approach called the Kemeny-Young method [22] used in electoral systems to identify the most popular choices in an election. Given a candidate global ranking, any ordered pair in a partial ranking will be able to find its match in the global ranking. An agreement between a PI's order of the pair and the order in the global ranking will contribute one to the total count and a disagreement will not contribute. CI is normalized by the number of all votes on pairwise orders given by PIs, and calculates the proportions of pairs among all sources (PIs) that are ranked agreeable to the candidate ranking. As introduced in the simulation setup subsection, CI is used to quantify the quality of rankings, and has been used in the standard performance measure for model assessment in survival analysis. Higher CI means better agreement between the final ranking and the ranks provided by the applicants. To be more specific, we describe the calculation of CI values in our method as follows. Given a global ranking of $n_p$ proposals, each applicant will review $n_r$ proposals. The ranking of $n_r$ proposals by each applicant will give relative ranks for $n_r (n_r$-1)/2 proposal pairs. The total number of pairs from all applicants is $n_p *n_r (n_r$-1)/2, with the possibility that some pairs may occur more than once. We compare the relative ranks of the $n_p *n_r (n_r$-1)/2 pairs provided by the applicants with the relative ranks of the corresponding proposal pairs in a given global ranking. The fractions of the pairs that both agree is the CI value for the global ranking.

While CI as a criterion is well founded, to obtain a global ranking with the highest CI value is not trivial. We implemented an accelerated simulated annealing algorithm to search a global ranking with the optimal CI. Parameters were tuned to fit different $n_p$ and $n_r$ to avoid local maxima. Comparison between the global ranking of the algorithm and the ranking obtained manually by brute force for small examples showed that they are consistent to each other. The algorithm is described in detail in the appendix.

## III. RESULTS
### 1) Simulation results

We use CI and $T_{0.2}$ as the performance measurements for most of the analyses.

We begin with the study of the effects of parameters $n_p$ and $n_r$ to the ranking accuracies. This will help us to determine the size of the PI pool and the number of proposals each PI should review. We are also interested in studying how PI behaviors such as $E_r$, $B_r$ and the true score variations $SD_s$ will affect the ranking accuracies, in order to test the robustness of the DPR approach. When studying the effects of these parameters, we use MBC to obtain the global ranking. Unless otherwise stated, the values of $B_r$, $E_r$ and $SD_s$ are kept to 10, 10 and 20 throughout the paper, in the cases they are not serving as changing variables of interest.

**Number of PIs per review group, $n_p$.** Figure 1 shows the effect of $n_p$ on CI and $T_{0.2}$. Interestingly, it appears that number of reviewer pool affects the performance measured by CI more than that measured by $T_{0.2}$. The overall magnitudes of effect for both CI and $T_{0.2}$ are modest. In the rest of the paper we set 40 as the upper limit of $n_p$.

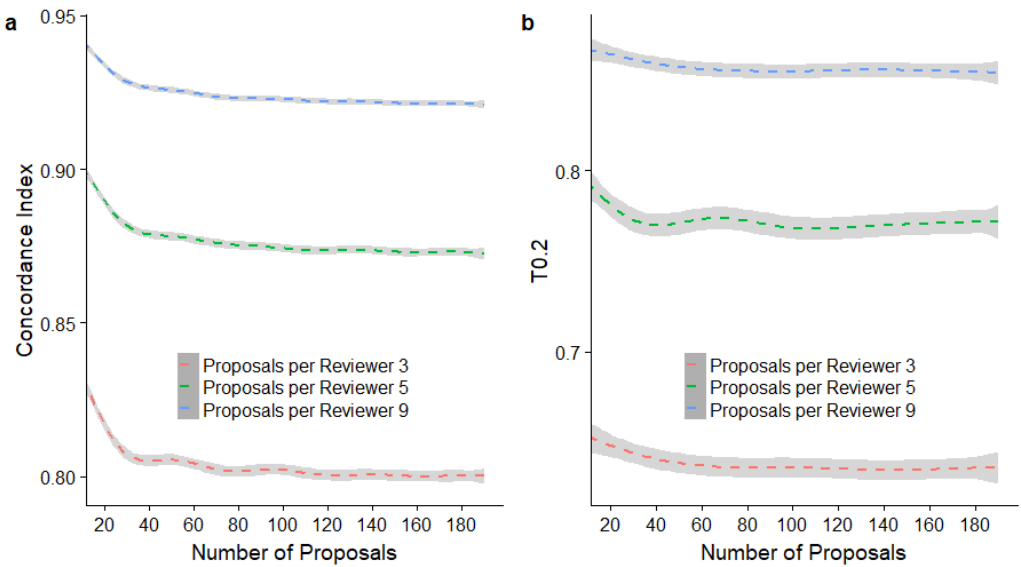

**Figure 1. Effect of number of proposals in a review pool.** (a) Performance measured by concordance index (CI) for three different numbers of proposals per reviewer, 3, 5, and 9; (b) Performance measured by $T_{0.2}$ for three different numbers of proposals per reviewer, 3, 5, and 9. $E_r = 10$ and $SD_s = 20$. The widths of the curves show their 99.9% confidence intervals. 1000 simulation replicates were carried out for each combination of parameters.

**Number of proposals assigned to each reviewer, $n_r$.** Figure 2 shows the change of CI (Figure 2a) and $T_{0.2}$ (Figure 2b) when $n_r$ increases. Clearly, with higher number of proposals each reviewer evaluate, the better accuracy of the global rankings will be. The scaling behaviors of the curves are similar with varying values of the review error, $E_r$, and the result indicates that 5-9 proposals for each reviewer will likely strike a balance between satisfactory review accuracy and workload. There is no need to keep increasing the number of reviews for each reviewer after certain values as the gain would diminish significantly. Even with reduced values of $n_r$, the accuracies are still acceptable depending on the expected accuracy level one would like to achieve and the quality of reviews.

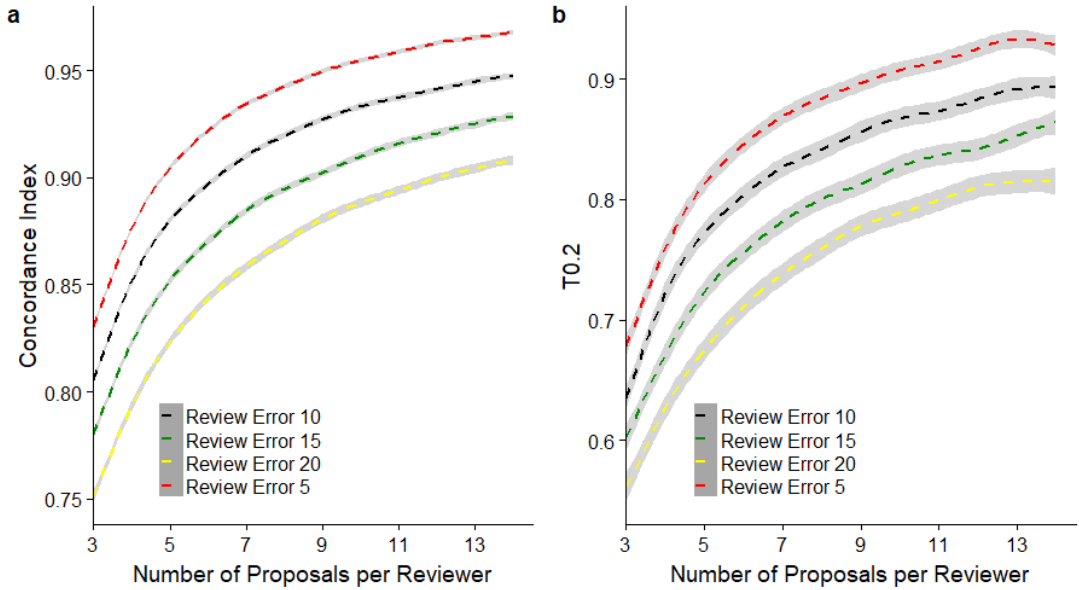

**Figure 2. Effect of number of proposals per reviewers to the performance. (a)** Concordance index for four different review errors; **(b)** $T_{0.2}$ for four different review errors. The widths of the curves show their 99.9% confidence intervals. 1000 simulation replicates were carried out for each combination of parameters.

**Bias of reviewers, $B_r$.** In principle, when a reviewer has the same bias towards all the proposals he/she evaluates, the rank of the proposals will not be affected by such bias. This can be seen clearly from our results (Figure not shown).

**Average error of the reviewer's scores, $E_r$.** Figure 3 shows a linear, negative relationship between $E_r$ and CI (Figure 3a), and between $E_r$ and $T_{0.2}$ (Figure 3b). As expected, smaller $E_r$ will give better global ranking accuracy.

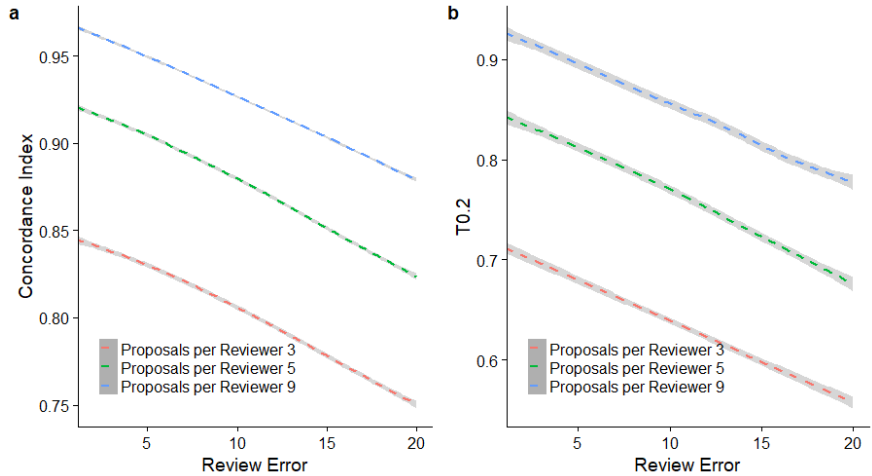

**Figure 3. Effect of average review error of reviewers on the performance.** (a) Concordance index for three different numbers of proposals per reviewers. (b) $T_{0.2}$ for three different numbers of proposals per reviewer. The widths of the curves show their 99.9% confidence intervals. 1000 simulation replicates were carried out for each combination of parameters.

**Standard deviation of true scores, $SD_s$.** Intuitively, the lower the $SD_s$ value, the harder it is to distinguish proposals with similar true scores. Figure 4 shows the effect of $SD_s$ on CI and $T_{0.2}$. It can be seen that there is significant increase in CI at small values of $SD_s$, followed by much slower increase of CI at larger values. The "turning point" is likely correlated with reviewers' review quality in terms of the error of their scores, $E_r$. When $E_r$ is smaller, reviewers are modeled as having higher abilities in distinguishing proposals of similar merit (true scores). From both Figures 4a and 4b, we can see how $E_r$ and $SD_s$ together affect the ranking accuracies. We can also find the values of $E_r$ and $SD_s$ to maintain the ranking accuracy above certain level. This can be useful when actual values of such parameters can be estimated from previous data or from empirical knowledge. The significant increase of CI and $T_{0.2}$ at smaller values of $SD_s$ is also interesting in the sense that the standard deviation of scores can be estimated from real data and rubrics can be made to better separate the proposals to increase the standard deviation of scores. This result provides clear support for establishing better evaluation guidelines and also some insight into how much increase in the standard deviation of scores should be made to reach a sufficient level of global ranking quality.

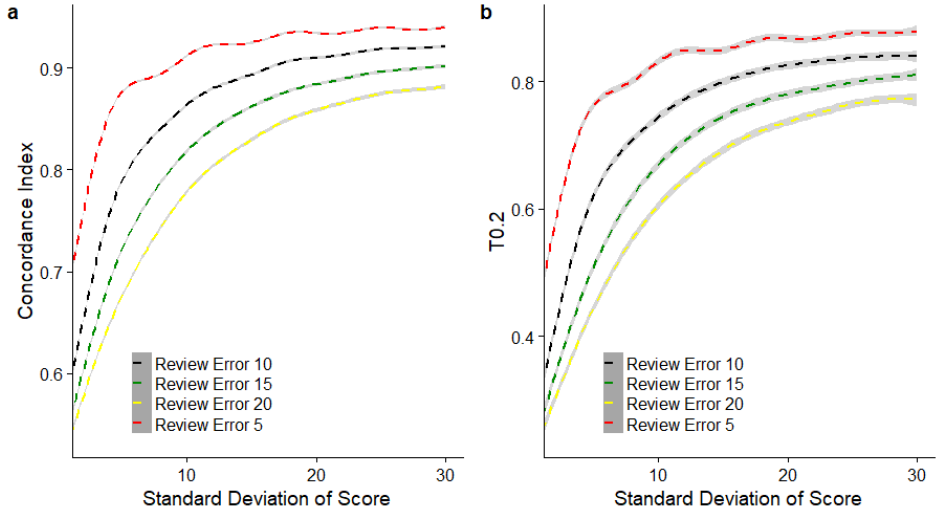

**Figure 4. Effect of standard deviation of true scores on the performance. (a) Concordance index for four different numbers of review errors; (b) $T_{0.2}$ for four different review errors. The widths of the curves show their 99.9% confidence intervals. 1000 simulation replicates were carried out for each combination of parameters.**

*2) Comparison between CIGR and MBC*

We compare the performance of CIGR and MBC at different values of average errors of reviewers (Figure 5). We find, somewhat to our surprise, that CIGR has an advantage over MBC when the quality of PI's reviews is relatively high, but MBC is more robust as the quality declines. The corresponding p-values are: 3.25e-47, 9.07e-05, 0.0080, and 8.71e-12, for review errors 5, 10, 15, and 20, respectively. The comparison when using filtering accuracy $T_{0.2}$ as the performance measure gives us similar results.

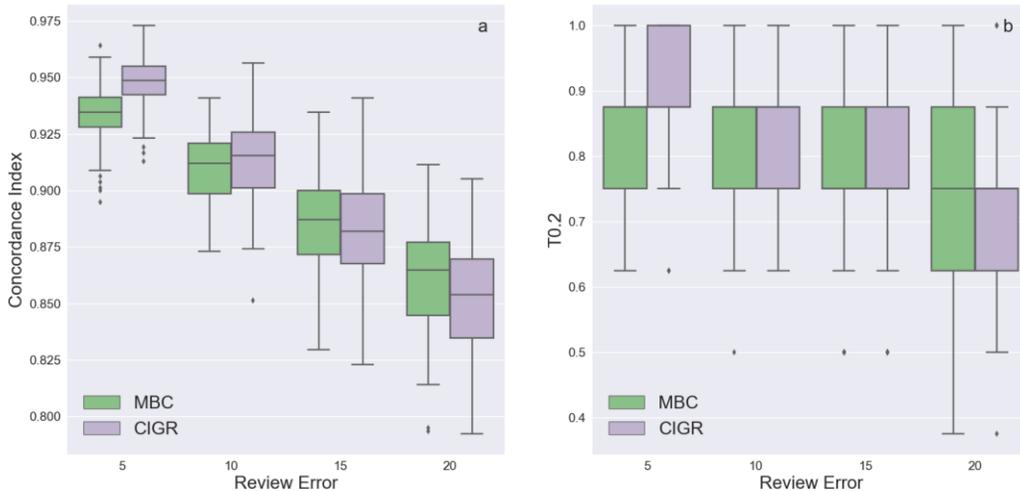

**Figure 5. The comparison between CIGR and MBC for four different review errors (5, 10, 15, and 20). CI: concordance index. The p-values are 3.25e-47, 9.07e-05, 0.0080, and 8.71e-12, respectively. 200 simulation replicates were carried out for each combination of parameters.**

*3) Boundary betwenn MBC-superior and CIGR-superior regions*

Since the performance of MBC and CIGR depends on the average error of the reviewer's scores, $E_r$, it will be interesting to find the occasions when MBC (or CIGR) has better performance. Because $E_r$ and $SD_s$ together determine the accuracy of the estimated global rankings, we identify the boundary on the $SD_s$-$E_r$ plot separating the MBC superior region from the CIGR superior region. Practitioners will then be able to use the plot as guidance to choose one of the two methods based on their knowledge on the review quality and the difficulty of the review task at hand.

We investigate a range of $SD_s$ values from 1 to 30, and locate the corresponding values of $E_r$ (the intersections) to fit a regression line with 99% confidence interval. This boundary shows in which regions CIGR outperforms MBC (Figure 6).

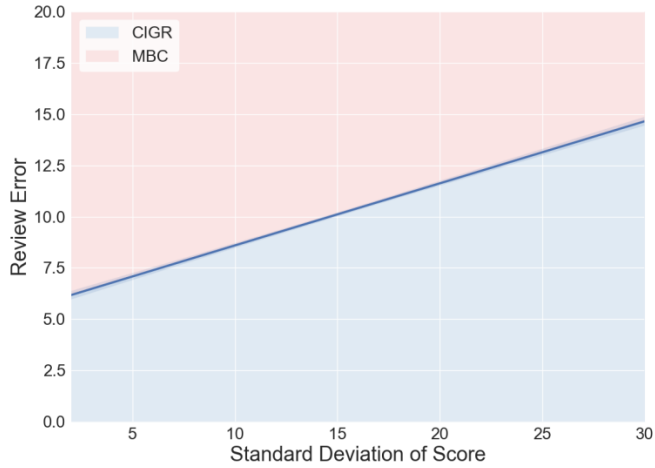

**Figure 6. The boundary between the CIGR-superior and MBC-superior regions on the $SD_s$-$E_r$ plot. 100 simulation replicates were carried out for each combination of parameters.**

*4) A balanced review assignment*

In this study, we also design a new method to assign proposals to the PIs so that the number of distinct pairs of proposals assigned to the same PI distribute uniformly. The idea is that we would like each proposal to be compared to as many as other proposals as possible. The details of the algorithm, called balanced assignment, is given in appendix. We can see from Figure 7 that balanced assignment can help both MBC and CIGR to significantly improve their ranking accuracies, measured by either CI (Figure 7a) or $T_{0.2}$ (Figure 7b).

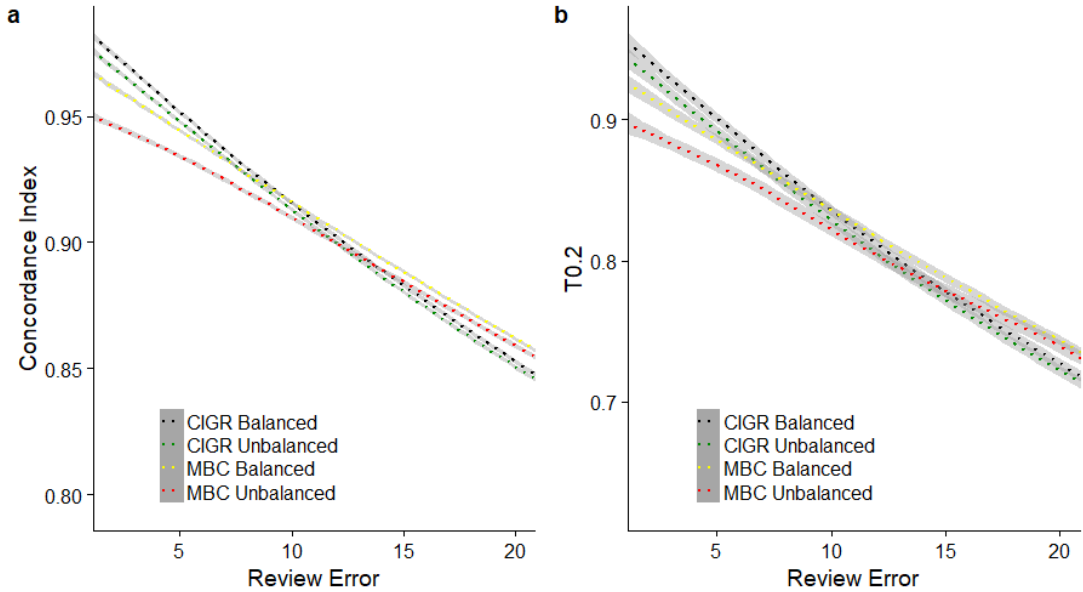

**Figure 7.** CI and $T_{0.2}$ comparison for MBC and CIGR for balanced and unbalanced assignments as a function of review error. The number of applicants is 40 and the number of reviews from each applicant is 7. The curves are calculated by taking the average CI or top 20% over 1000 simulations for uniformly spaced points on the x-axis. Balancing improves the performance of both MBC and CIGR at all reliability values. The improvement in MBC for small values of review errors is particularly strong. The widths of the curves show their 99.9% confidence intervals.

## IV. CONCLUSIONS AND DISCUSSIONS

In this work, we perform simulation studies for a new distributed peer review (DPR) approach. The new DPR approach employs the applicants themselves to review proposals submitted by the other applicants. Each applicant reviews a relatively small number of proposals and submits a rank for the proposals they review. A global ranking is then derived from the individual rankings. We examine the effect of several key parameters for the decision making when using a DPR approach. Our simulation studies can provide useful guidelines when adopting a DPR approach in real peer review settings.

Merrifield and Saari [13] also proposed to use the DPR approach for reviewing scientific proposals, and performed some simulations to evaluate the effectiveness of this method. In their paper, the uncertainty of reviews was modeled by assuming that every reviewer was fundamentally unable to distinguish between applications within $\Delta n = 10$ places of each other in the true ranks, but otherwise can rank each application fairly against its competitors. The cutoff value of 10 is somewhat arbitrary. In addition, they also assumed that all the reviewers had the same review quality, which may not be satisfied in the real situation. Our study simulates a more realistic process and, as a result, the conclusions should be more applicable to real review tasks.

We also propose a new method Concordance Index-based Global Ranking (CIGR) for inferring the global ranking from individual rankings. The main purpose of CIGR is to maximize the proportion of matching pairs in a global ranking to the pairwise orders given by all reviewers – i.e., a larger proportion of matching corresponds to a larger value of CI and thus a better global ranking. An effective MCMC-based algorithm is developed to search for near optimal global rankings with respect to CI, and a stable and high quality global ranking is computed by aggregating over a collection of near optimal rankings. We find that CIGR outperforms MBC in cases where review qualities are relatively high.

Furthermore by using CIGR we can modify the current DPR approach to a multi-stage review process to potentially achieve even better review quality. This multi-stage process can be constructed in the following way. In the first step, each reviewer will be only assigned a relatively small number of proposals to review. The global rankings of all the proposals generated in this first step will be used to assign proposals to applicants in the next step. The idea is that we will assign proposals with similar rank to the same applicants with higher probability. With this method proposals with similar quality will be more likely to be grouped together and have head-to-head comparisons in the second step, yielding a better final global ranking. Note that the MBC-based criterion will not work in this modified procedure, because all the proposals except the top- and bottom-ranked ones will have similar MBC values.

We can also eliminate a proportion of proposals step-by-step in this multi-stage process. For example, we may filter out the lower 50% of the proposals after we generate the global ranking in the first step because these proposals may safely be considered uncompetitive. This elimination step can be repeated until a desired number of proposals is selected. In this case, one may handle a very large number of proposals in a single review task and generate higher quality reviews while substantially lowering the workload of each reviewer.

We further design an algorithm to create a balanced assignment, which assigns proposals to reviewers by making the number of times each pair of proposals has a head to head comparison from the same reviewer as uniform as possible. This balanced assignment technique helps to improve the performance of both the MBC and CIGR methods.

In summary we believe the DPR approach is a very promising new method for peer review which may resolve some of the long-standing problems in the current peer review process. We encourage the scientific community to perform additional research and test such an approach when possible. We hope it will be adopted when enough evidence on its effectiveness is collected.

**Competing Interests**

Jinfeng Zhang and Tingting Zhao are founders of Pevals LLC.

**Author contributions**

A.S. designed and implemented the core algorithms and helped implement and parallelize the simulation pipeline. J.Q. implemented and helped design the simulation pipeline, performed exploratory analysis, produced the figures and graphics, and suggested improvements to the algorithms. J.Z. contributed to the design and implementation of the simulation pipeline and algorithms. M.T., T.Z., X.P., and J.Z. designed the study. A.S., J.Q., and J.Z. wrote the manuscript. M.T., and T.Z revised the manuscript. All authors reviewed the manuscript.

APPENDIX

SIMULATED ANNEALING ALGORITHM FOR CIGR

If the $n$ PIs have each been assigned $m$ proposals to rank, then the $RCM$ matrix corresponding to the totality of their rankings is given by

$$RCM_{ij} = \text{The number of reviewers who reviewed proposals } i \text{ and } j \text{ and ranked } i > j$$

Given a candidate ranking, $\boldsymbol{r} = (r_1, r_2, \ldots, r_n)$, the cost function associated to $r$ is

$$C(\boldsymbol{r}) = \sum_{i=1}^{n} \sum_{j=i+1}^{n} \left[ RCM_{r_i r_j} < RCM_{r_j r_i} \right]$$

Given a pair $(i,j)$ with $i < j$, $(r_i, r_j)$ agrees with the $RCM$ matrix if $RCM_{r_i r_j} \geq RCM_{r_j r_i}$, otherwise it disagrees with the $RCM$ matrix. The cost function counts the number of pairs in the ranking $\boldsymbol{r}$ that disagree with the $RCM$ matrix.

We employ a simulated annealing algorithm to find global rankings $\boldsymbol{r}$ with near minimal cost. Starting with an initial ranking $\boldsymbol{r_0}$, an initial temperature $T_0 \geq 0$, and a cooling parameter $\beta$ with $0 \leq \beta \leq 1$, a matrix is used to keep track of which pairs in the ranking disagree with the $RCM$ matrix. At iteration $k$, a candidate ranking is proposed by switching the order of a pair in the current ranking $\boldsymbol{r_k}$ that disagrees with the $RCM$ matrix. If the candidate ranking $\boldsymbol{r}^*$ lowers the cost function, this candidate is accepted and replaces the current ranking. Otherwise, the candidate ranking is accepted with probability $e^{-(C(\boldsymbol{r}^*)-C(\boldsymbol{r_k}))/T_k}$. If the candidate ranking is not accepted, the current ranking remains the same in the next iterations. $T_{k+1}$ is then set to $\beta T_k$, making it less likely that a suboptimal candidate is selected in the next iteration. The algorithm keeps track of the best ranking that has been seen so far and upon termination, returns this ranking.

Through experiment, we have observed that the CIGR cost function may have many locally optimal kinetic traps. To explore more of the search space, the algorithm employs restarts. The algorithm maintains a set of all global rankings within a threshold ε of the best ranking found so far and keeps track of the number of iterations $j$ that have passed since a candidate ranking has been accepted. The algorithm also keeps track of the number of pairs Θ in the current ranking that disagree with the $RCM$ matrix. Θ is equal to the number of potential candidate rankings during an iteration. The algorithm employs a patience parameter $\rho$. When $\frac{j}{\Theta}$ exceeds $\rho$, the algorithm restarts by choosing a new initial ranking from the maintained set of near optimal rankings and resetting the temperature to its initial value. The algorithm terminates after a set number of restarts or a set number of total iterations. At the end, a global ranking is generated by applying MBC to the set of all near optimal rankings that the algorithm has found. We have found that this aggregate of near optimal rankings tends to achieve superior performance to an exact minimizer, though this can only be demonstrated in simulations where the number of proposals is sufficiently small to make computation of an exact solution feasible.

## OPTIMAL BALANCING OF ASSIGNMENTS

Given $n$ principal investigators $p_1, p_2, \ldots, p_n$ who each submit proposals, we seek an assignment of $m$ proposals to each PI such that no PI is assigned their own proposal. More generally, to each PI, $p_i$, we can associate a set of proposals $C_i \subseteq \{1, \ldots, n\}$ that $p_i$ is not permitted to review. We shall call these the constraints of an assignment problem. Note that for some sets of constraints a valid assignment may not exist. We are interested in assignments such that the $\binom{n}{2}$ pairs of proposals are as uniformly distributed among the $n$ reviewers as possible.

Our strategy is to first generate an assignment randomly and then employ stochastic optimization to iteratively balance this assignment. We use the Shannon Entropy of an assignment as a measure of its balance. Each of the $n$ reviewers is assigned $m$ proposals, and thus $\binom{m}{2}$ pairs of proposals. $n\binom{m}{2}$ pairs will then be compared within an assignment, with the possibility of duplicates. These $n\binom{m}{2}$ pairs are to be distributed among the $\binom{n}{2}$ possible pairs of proposals. In a perfectly balanced assignment, each of these $\binom{n}{2}$ pairs would be assigned to $\frac{m(m-1)}{n-1}$ of the reviewers, with the understanding that this isn't possible for most values of $n$ and $m$. Given an assignment A, for each pair of proposals $(i,j)$, let $\alpha_{ij}$ be the number of principal investigators who have been assigned both $i$ and $j$. Then the Shannon entropy of the assignment A is given by

$$H(A) = \sum_{i=1}^{n} \sum_{j=i+1}^{n} -\frac{\alpha_{ij}}{n\binom{m}{2}} \log \frac{\alpha_{ij}}{n\binom{m}{2}}$$

Where it is assumed by convention that a term of the sum is equal to 0 if $\alpha_{ij} = 0$.

As is known from information theory, $H(A)$ will achieve its maximum value for a perfectly balanced assignment, and gives a good overall measure for the balance of an assignment. Like the case of computing a Concordance Index based Global Ranking, we will employ a simulated annealing strategy to prevent the algorithm from getting stuck in kinetic traps, which in this case can result from the constraints. To develop such an algorithm, we need a means of making small modifications to an assignment to generate candidate assignments, which can then be accepted or rejected. Here, assignments are modified by having a pair of principal investigators trade proposals such that the new assignment still satisfies the constraints.

Given a pair of principal investigators $(p_i, p_j)$ who have been assigned the proposals $r_i$ and $r_j$, $r_i$ and $r_j$ are a tradeable pair if $r_i \neq r_j$, $p_i$ is allowed to review $r_j$, and $p_j$ is allowed to review $r_i$. At each iteration, a tradeable pair is selected uniformly at random. A candidate assignment is generated by making the trade and the difference in entropy is calculated between the new and old assignment. If the entropy increases, the candidate is automatically accepted. Otherwise it is accepted with probability $e^{-(H(A_k) - H(A^*))/T_k}$. The temperature is multiplied by the cooling parameter and the algorithm continues until the stopping criterion is met.